\documentclass[sigconf,natbib=true]{acmart}
\usepackage{array}
\usepackage{multirow}
\setcopyright{acmlicensed}
\copyrightyear{2018}
\acmYear{2018}
\acmDOI{XXXXXXX.XXXXXXX}
\acmConference[Conference acronym 'XX]{Make sure to enter the correct
  conference title from your rights confirmation email}{June 03--05,
  2018}{Woodstock, NY}
\acmISBN{978-1-4503-XXXX-X/2018/06}

\begin{document}

\title[Your plan may succeed, but what about failure?]{Your plan may succeed, but what about failure? Investigating how people use ChatGPT for long-term life task planning}

\author{Ben Wang}
\email{benw@ou.edu}
\orcid{0000-0001-8612-1185}
\affiliation{%
  \institution{The University of Oklahoma}
  \city{Norman}
  \state{Oklahoma}
  \country{USA}
}

\author{Jiqun Liu}
\email{jiqunliu@ou.edu}
\affiliation{%
  \institution{The University of Oklahoma}
  \city{Norman}
  \state{Oklahoma}
  \country{USA}
}

\renewcommand{\shortauthors}{}

\begin{abstract}
Long-term life task planning is inherently complex and uncertain, yet little is known about how emerging AI systems support this process. This study investigates how people use ChatGPT for such planning tasks, focusing on user practices, uncertainties, and perceptions of AI assistance. We conducted an interview study with 14 participants who engaged in long-term planning activities using ChatGPT, combining analysis of their prompts and interview responses. The task topics across diverse domains, including personal well-being, event planning, and professional learning, along with prompts to initiate, refine, and contextualize plans. ChatGPT helped structure complex goals into manageable steps, generate ideas, and sustain motivation, serving as a reflective partner. Yet its outputs were often generic or idealized, lacking personalization, contextual realism, and adaptability, requiring users to actively adapt and verify results. Participants expressed a need for AI systems that provide adaptive and trustworthy guidance while acknowledging uncertainty and potential failure in long-term planning. Our findings show how AI supports long-term life task planning under evolving uncertainty and highlight design implications for systems that are adaptive, uncertainty-aware, and capable of supporting long-term planning as an evolving human–AI collaboration.
\end{abstract}



\maketitle

\section{Introduction}

Long-term life tasks, such as health management, financial planning, or education, are extended, multi-step activities unfolding over weeks, months, or years. Unlike short tasks that can be completed quickly and have clear boundaries, long-term tasks require sustained attention and iterative decision-making to coordinate multiple evolving steps \citep{norman_attention_1986, card_psychology_2008, allen_getting_2015}. Effective planning for such tasks is essential for achieving meaningful long-term goals in personal development, well-being, and everyday life.

A central challenge in long-term task planning is uncertainty, a condition of limited knowledge or predictability about future states and actions \citep{knight_risk_1921, simon_behavioral_1955}. In task management contexts, for example, two common forms of uncertainty are outcome uncertainty (i.e., not knowing what results will follow from one's efforts) and action uncertainty (i.e., not knowing which steps are needed or most effective to achieve a goal) \citep{maes_relationship_2022}. Such uncertainty increases cognitive load and often undermines motivation in planning and task processes \citep{abbott_understanding_2005}. While individuals employ planning and task management strategies to cope with uncertainty, human cognitive limitations make it difficult to maintain coherence and motivation across extended timescales \citep{kahneman_maps_2003, tversky_advances_2000}.

\begin{figure*}
  \includegraphics[width=0.95\textwidth]{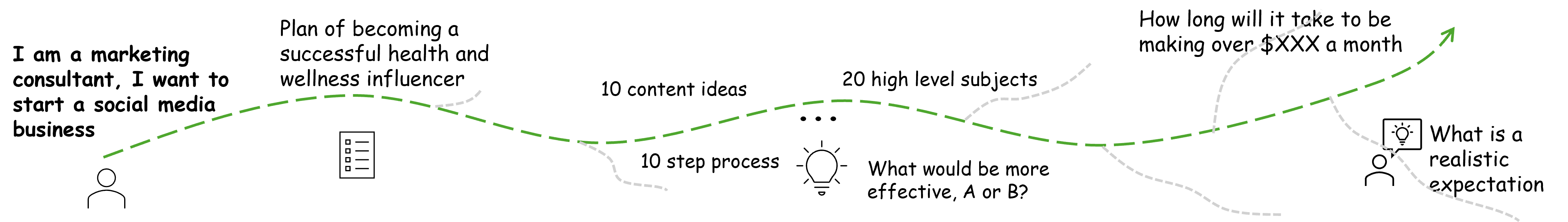}
  \caption{An illustration of the planning process of one participant. The main green line represents the path planned with ChatGPT. The gray branches represent hidden paths leading to different results or potential failure.}
  \label{fig:path}
\end{figure*}

Existing tools such as search engines, to-do lists, and calendars support specific aspects of tasks, such as information seeking, time scheduling, or task tracking \citep{bellotti_what_2004, shah_envisioning_2024}. However, they are not designed to integrate these components into a coherent long-term planning process. As a result, people invest considerable effort to interpret and organize information, and monitor progress on their own, facing challenges in maintaining coherence, motivation, and adaptability over time. Recent advancements in large language models (LLMs) have opened new possibilities for human–AI collaboration in task planning. LLM-based chatbots, such as ChatGPT, enable open-ended, multi-turn interactions, offering users personalized advice, information synthesis, and creative ideation \citep{brown_language_2020, ouyang_training_2022}. Usage patterns show that people increasingly rely on ChatGPT for complex, multi-step tasks such as practical guidance, information seeking, and content generation \citep{chatterji_how_2025, skjuve_user_2023}. OpenAI has also introduced ChatGPT's learning mode, which supports step-by-step study and long-term skill development \citep{openai_introducing_2025}.
These developments suggest that users and developers are increasingly exploring LLMs as partners for complex and long-term activities.

However, research in human–AI interaction has not yet caught up with how people are increasingly using LLMs for complex activities \citep{xu_ai_2020, gomez_human-ai_2025}. Existing studies on LLM collaboration on long-term tasks have explored topics such as travel planning \citep{xie_travelplanner_2024}, research assistance \cite{tang_ai-researcher_2025, schmidgall_agent_2025}, team collaboration \citep{xu_theagentcompany_2025}, and even social interaction \citep{piao_agentsociety_2025}. Yet most of these studies rely on simulated tasks rather than investigating real user experiences. Some emerging works with real users focus on programming assistance \citep{feng_coprompt_2024} or educational contexts \citep{urban_chatgpt_2024}. However, little is known about how real users experience LLMs as collaborators in planning and managing long-term life tasks, how these systems affect uncertainty in tasks, and what opportunities they create for future task-support systems.

This paper addresses this gap through an interview study of how people use ChatGPT for long-term life task planning. We recruited participants to engage in self-chosen planning tasks with ChatGPT, covering topics of personal and well-being, activity and event planning, and professional development and learning, followed by semi-structured interviews. Figure \ref{fig:path} illustrates one planning path from the participants. By analyzing both their prompts and interview reflections, we capture not only how people interacted with an LLM but also how they perceived uncertainty, usefulness and limitations in AI-assisted long-term life task planning.

Our findings show that participants engaged with ChatGPT through diverse prompting strategies that varied across task domains, including personal and well-being, activity and event, and professional development and learning tasks. They encountered two main sources of uncertainty, task-inherent (e.g., where to start, feasibility) and AI-introduced (e.g., contextual fit, credibility), which influenced both emotion and behavior. ChatGPT helped structure goals and encourage reflection, but general or overconfident advice sometimes amplified uncertainty. Overall, participants viewed ChatGPT as a useful yet limited planning partner that provided scaffolding, motivational support, and reflection, while lacking adaptability and personalization.

This study advances research on human–AI collaboration by foregrounding long-term life tasks as a distinct and meaningful context for studying LLM usages. It contributes (1) empirical insight into how users actually engage with conversational AI for long-term life task planning; (2) key uncertainties users encounter when planning with AI and the related impacts; and (3) design implications for AI systems that serve not only as scaffolds for success but also as adaptive, reflective partners that help users navigate uncertainty, challenges, and potential failure in long-term goals.

\section{Background}

Long-term life tasks represent the complex and sustained goals that individuals pursue across different stages of life, such as managing finances, sustaining health, advancing education, or building a career \citep{havighurst_human_1953, hassan_conceptual_2021, fruh_obesity_2017, young_factors_2019, pinheiro_take_2017}. These tasks shape identity and fulfillment but are inherently challenging due to their open-ended nature, uncertainty, and need for continuous adjustment \citep{grant_impact_2003, lieder_cognitive_2019, woolley_immediate_2017}. As such, effective planning and decision-making are critical, yet existing tools and approaches only partially support these needs. This section reviews three key strands of research, including task management, information access systems, and the potential of LLMs, to situate the gap addressed by this study.

\subsection{Task management approaches to planning}
Task management provides structured methods for organizing and completing goals, often breaking complex challenges into manageable steps \citep{allen_getting_2015, jones_planning_2006}. Goal setting helps sustain motivation and direct effort by linking immediate actions with broader objectives \citep{locke_building_2002, hochli_how_2018}. Planning decomposes these goals into concrete ``when, where, and how'' strategies that anticipate obstacles and support self-regulation over time \citep{sniehotta_action_2005, neenan_cognitive_2020}. Planning also involves managing uncertainty, which influences both decision-making and emotion \citep{abbott_understanding_2005, gu_uncertainty_2020}.

A wide range of systems have been developed to support task management. Personal Task Management (PTM) tools such as to-do lists and scheduling apps help externalize goals and priorities \citep{bellotti_what_2004, jones_personal_2017}. More advanced systems introduce structured pathways for complex tasks (e.g., task tours) \citep{hassan_task_2012}, predictive workload management \citep{takeuchi_task-management_2013, mita_early_2017}, and proactive intelligent assistants that anticipate user needs \citep{yorke-smith_design_2012, myers_intelligent_2007}. Recent work has highlighted integrating AI or LLM capabilities into task management, emphasizing their potential to complement human abilities \citep{barcaui_who_2023, white_navigating_2023}.

However, most existing systems focus on professional or collaborative contexts rather than personal long-term life tasks. They primarily assist users in structuring and scheduling plans, while the ongoing effort of adapting, maintaining motivation, and reflecting on progress still largely depends on humans. Although recent studies highlight the potential of AI-assisted task management, empirical evidence on how such systems can support personal long-term planning remains limited.

\subsection{Information seeking and access systems}
Complex tasks often involve subtasks that require acquiring and interpreting new information. Within this broader process, information tasks refer to the activities users undertake to locate, evaluate, and apply information in support of their overarching goals \citep{bystrom_conceptual_2005, vakkari_task-based_2003}. Information seeking is therefore an integral part of task completion, and it introduces its own forms of uncertainty, as users may face knowledge gaps and ambiguous objectives, which affect emotion and task completion \citep{belkin_anomalous_1980, kuhlthau_inside_1991, keller_uncertainty_2020, azzopardi_cognitive_2021}. Research in Interactive Information Retrieval (IIR) examines how users interact with search systems to manage such information uncertainty and perceive results in user-centered, multi-query interactions \citep{liu_interactive_2019, petrelli_role_2008}. Unlike traditional system-centric evaluation, IIR emphasizes user experience measures such as usefulness, satisfaction, and task effectiveness \citep{cole_usefulness_2009, almaskari_review_2010}.

Building on the user-centered vision of IIR, adaptive and proactive information systems were developed to anticipate user needs and provide real-time support throughout information seeking \citep{bhatia_proactive_2016, liu_state-aware_2021}. These systems span multiple paradigms, including search engines, conversational assistants, and recommendation systems, which use behavioral signals and contextual cues to predict users' intent, struggles, and affective states \citep{jannach_survey_2022, alonso_conversational_2024}. By modeling such factors, they aim to reduce information uncertainty and help users formulate and refine information tasks more effectively. Despite these emphases, most adaptive systems remain limited to information interactions rather than supporting broader task management. Recent developments in LLMs have renewed interest in integrating conversational interactions and world knowledge into these systems, enabling more flexible and context-aware support for complex, evolving tasks \citep{deng_towards_2024, shah_envisioning_2024}.

\subsection{Potential and challenges of LLMs for long-term life task planning}
Compared with traditional task management and information access systems, LLMs introduce a new paradigm for long-term task support. They enable multi-turn, natural-language interaction, allowing users to iteratively refine goals, clarify context, and adapt plans as needs evolve \citep{shah_taking_2023, liao_proactive_2023}. LLMs can act as interactive partners that integrate contextual information and world knowledge to address more complex, open-ended queries \citep{liu_pre-train_2023, skjuve_user_2023}. Beyond information access, they demonstrate emerging reasoning and planning abilities, such as decomposing tasks into subtasks and applying strategies like Chain-of-Thought reasoning \citep{wei_chain--thought_2022, yao_react_2022}, which enable them to function as agents for task support and automation \citep{huang_understanding_2024, wu_autogen_2023}.

However, these opportunities come with significant challenges. Users often struggle with prompt formulation, interpretation, and overreliance when facing uncertainty \citep{zamfirescu-pereira_why_2023, salimzadeh_dealing_2024}. Human-centered AI research highlights the importance of transparency, explainability, and user control for trust and effective collaboration \citep{amershi_guidelines_2019, hu_chatgpt-cps_2024, kim_help_2023}. Existing work demonstrates that LLMs can assist with long-term planning in simulated environments, such as travel planning \citep{xie_travelplanner_2024}, research assistance \cite{tang_ai-researcher_2025, schmidgall_agent_2025}, team collaboration \citep{xu_theagentcompany_2025}, and even social interaction \citep{piao_agentsociety_2025}, as well as real-world applications for programming, writing, and education \citep{feng_coprompt_2024, wang_task_2024, urban_chatgpt_2024}. Yet most remain domain-specific, short-term, or lack insight into how real users collaborate with LLMs in personal, open-ended, and uncertain long-term life tasks. Understanding these everyday practices and challenges is essential for designing AI systems that can effectively support long-term planning under uncertainty.

\section{Research Questions}

This study introduces \textit{long-term life tasks} as a novel task type for human-AI collaboration. People increasingly use LLMs for complex work. However, there remains limited understanding of how they are used for long-term life task planning, what kinds of uncertainty users experience in this process, and how such systems should be designed to provide effective support. Accordingly, this research investigates the following questions:

\begin{itemize}
    \item[\textbf{RQ1}:] What types of long-term tasks do users choose to plan with the support of generative AI?  
    \item[\textbf{RQ2}:] What kinds of uncertainty do users experience during AI-assisted long-term task planning?  
    \item[\textbf{RQ3}:] How do users perceive the role of generative AI in helping them manage task planning and uncertainty?  
\end{itemize}

\section{Methodology}
This study employs a qualitative research design to explore how AI users cope with uncertainty in long-term life task planning. The research consisted of two main components: a warm-up study and an in-depth semi-structured interview. The warm-up activity provided participants with a hands-on experience in AI-assisted task planning. Using the long-term goal defined during the warm-up as a reference, the interview gathered participants' feedback on their experience and and perspectives on human–AI collaboration in long-term task planning.

\subsection{Participant recruitment}
As an exploratory study, we conducted in-depth interviews with a small number of participants to examine their contexts, motivations, and challenges in AI-assisted long-term planning. This study recruited 14 participants (aged 20 to over 80) with diverse educational and professional backgrounds, including students, engineers, managers, educators, and researchers. Seven participants were recruited via information schools' mailing lists and professional networks, and the remaining through the social media platform Reddit. Most participants reported using ChatGPT several times a day for both professional (e.g., study or work-related tasks) and personal purposes (e.g., health, daily schedule, or learning). Each study session (including the warm-up and interview) lasted around two hours. Each participant received \$30 USD as compensation. The study was approved by the university's Institutional Review Board.

\subsection{Interview procedure}
Participants first engaged in a warm-up activity designed to simulate AI-assisted long-term task planning using ChatGPT. The researcher collected demographic information and asking about participants' previous experiences of using ChatGPT or ideas for tasks where it might assist planning. The researcher and participant then selected a topic together for the warm-up activity. Examples of goals include pursuing higher education, starting a career, or adopting a healthier lifestyle. Topic complexity and timeframe varied, allowing exploration of diverse strategies and challenges.

Participants could choose to use their own prompts or the templates provided by the researcher. The prompt templates included three versions, each reflecting a step in generating and refining task plans. The first template generally included the participant's long-term goal and current status: ``\textit{I am [status], can you provide me with a [topic] plan}.'' The second template asked for details about a specific part of the plan generated in the first prompt: ``\textit{For [a specific part], can you provide me with a [step-by-step/weekly] plan}.'' The third template added more contextual information and asked ChatGPT to revise the plan: ``\textit{For [a specific part], I have [preference/consideration], can you revise the plan}.'' Participants using the provided templates received an initial task planning outline from ChatGPT and discussed any immediate impressions. They could then provide additional details about their situation or preferences to generate or revise content in the second and third prompts. Participants using their own prompts engaged with ChatGPT more flexibly, deciding what aspects of planning to explore. The researcher occasionally offered prompt templates if participants needed guidance. Participants used a free ChatGPT account provided by the researcher, with the memory function disabled to prevent influence from prior conversations.

Throughout this process, participants were encouraged to think aloud and discuss the usefulness of the AI-generated task decomposition. They reflected on how this AI-assisted planning compared to their usual non-AI methods, noting any insights or discrepancies. During the warm-up activity, the researcher asked participants interview questions about their experience after reviewing each ChatGPT response. As shown in Table \ref{tab:interview questions}, following the initial question on task topic selection, the interview consisted of four main sections exploring participants' experiences with AI-assisted long-term task planning. Part 1 examined participants' first impressions of ChatGPT's responses; Part 2 focused on experiences of uncertainty during planning; Part 3 addressed usefulness and limitations on specific task components; and Part 4 captured broader reflections, expectations, and suggestions for improvement.

\begin{table*}
\caption{Interview questions.}
\label{tab:interview questions}
\centering
\small
\begin{tabular}{l} \hline
\textbf{Part 0 Task planning topic~} \\ \hline
\begin{tabular}{@{\labelitemi\hspace{\dimexpr\labelsep+0.5\tabcolsep}}l@{}}\textit{Can you think of a specific task or goal in your life where AI could help you plan or organize things over a long period?}\end{tabular} \\ \hline
\textbf{Part 1 Initial impression} \\ \hline
\begin{tabular}{@{\labelitemi\hspace{\dimexpr\labelsep+0.5\tabcolsep}}l@{}}\textit{How do you think the plan is useful? $\bullet$ How do you feel ChatGPT understands your background when helping you with this task?}\\\textit{If you weren't using ChatGPT, how would you usually plan for this kind of task? How does that compare with using ChatGPT?}\\\textit{(If using the provided prompt templates) How do you feel about the way ChatGPT breaks down your task? What do you like or dislike?}\end{tabular} \\ \hline
\textbf{Part 2 Uncertainty in task planning} \\ \hline
\begin{tabular}{@{\labelitemi\hspace{\dimexpr\labelsep+0.5\tabcolsep}}l@{}}\textit{Which part of the AI-assisted plan made you feel uncertain? $\bullet$ How did that uncertainty affect your emotions or behavior?}\\\textit{How do you think AI is able to help reduce uncertainty or made it easier to deal with it in this planning process?}\\\textit{What kinds of uncertainty might you face if you were using non-AI methods instead?}\end{tabular} \\ \hline
\textbf{Part 3 Task planning details} \\ \hline
\begin{tabular}{@{\labelitemi\hspace{\dimexpr\labelsep+0.5\tabcolsep}}l@{}}\textit{Which part of the plan do you find most helpful? $\bullet$ Which part do you think ChatGPT struggles with when helping with your task planning?}\\\textit{Do you have any suggestions for how ChatGPT could improve in any specific part of the planning process?}\end{tabular} \\ \hline
\textbf{Part 4 Other aspects} \\ \hline
\begin{tabular}{@{\labelitemi\hspace{\dimexpr\labelsep+0.5\tabcolsep}}l@{}}\textit{Do you have a paid ChatGPT account? How do you think the paid version of ChatGPT performs better for this kind of task planning?}\\\textit{Would you prefer ChatGPT to give you multiple solutions to a problem, or just one clear path? Why?}\\\textit{How would you like ChatGPT to explain its recommendations or reasoning to you?}\\\textit{Do you think AI-assisted planning affects your ability to make your own decisions in any way?}\\\textit{What other changes or features would make AI-assisted task planning more useful or effective for you?}\end{tabular} \\ \hline
\end{tabular}
\end{table*}

These interview questions were adapted to each participant's task and interest to ensure relevance. For example, if a participant expresses interest in film making, an example question can be tailored to: \textit{How do you think AI can help you learn film making or create a structured plan for learning?} This approach makes the questions more concrete and personally meaningful. Additionally, technical concepts were adjusted based on participants' background knowledge. For instance, instead of referring to ``model version or size,'' the question can be framed in terms of ``free or paid ChatGPT account'' to make it more accessible to participants. Such phrasing helped maintain clarity and engagement.

For questions in part 2, the interviewer first introduced the concept of task uncertainty, explaining its two example forms: \textit{outcome uncertainty}, where individuals are unclear about what they want to achieve, and \textit{action uncertainty}, where they are unsure about the steps needed to reach their goals. Participants were encouraged to share their opinions on how uncertainty impacts long-term life task planning. The discussion would highlight that while uncertainty can motivate exploration and creativity, it can also increase anxiety. This section thus examined how different uncertainties emerge and how users perceive and manage them. 

All questions were integrated throughout the warm-up activity, with explanations and follow-ups as needed. After the activity, the interview continued to cover any remaining questions not previously discussed, especially for questions in parts 3 and 4.

\subsection{Interview data analysis}
We first organized participants' background information and prompt behaviors in tabular form to support cross-case comparison. We manually tagged prompts according to their functional roles (e.g., goal setting, task decomposition, refinement, contextualization) and mapped them to different phases of task planning. This facilitated the identification of prompting patterns and engagement styles.

To account for differences in prompting behavior, we have different focuses between template-based and self-generated prompts. For participants who mainly used templates, we examined how these prompts served as starting points to explore AI-assisted planning and stimulate reflection. For those who wrote prompts independently, we focused on their natural prompting patterns to understand how users might organically engage with ChatGPT in real-life planning scenarios. 

The interview data were analyzed using a reflexive thematic analysis approach \citep{braun_using_2006, braun_reflecting_2019}, combining deductive coding guided by the research questions with inductive identification of emergent themes. The analysis focused on participants' goals, strategies, uncertainties, and perceptions of AI-assisted long-term task planning.

For the interview transcripts, the lead researcher followed six phases of reflexive thematic analysis: (1) familiarization through repeated reading of transcripts; (2) identifying features related to the research questions; (3) generating initial codes both deductively and inductively; (4) organizing related codes into broader themes; (5) reviewing and refining these themes; and (6) defining and naming them, supported by representative quotes. While deductive codes reflected the study's analytical focus (e.g., task types, planning strategies, uncertainty types, perceptions of AI assistance), inductive coding captured unexpected concerns or novel strategies.

To enhance rigor, the researcher iteratively refined the codes and themes through repeated engagement with the data and incorporated feedback from a second researcher. The second researcher reviewed the coding framework and emergent themes, offering an external perspective for validation. In line with reflexive thematic analysis, we did not calculate inter-coder reliability metrics, as themes were treated as interpretive patterns rather than objective entities \citep{braun_using_2006, braun_reflecting_2019}. Instead, analytic consistency was ensured through systematic engagement with the data, iterative refinement, and transparent reporting by providing representative quotes for each theme in the findings. We selected quotes to illustrate both typical and divergent cases, ensuring that the reported themes reflected the range of participant experiences.

\section{Results}

\subsection{Planning task topics and methods}
For \textbf{RQ1}, we investigated the task topics and prompt behaviors. During the user activity in this study, participants engaged in planning tasks spanning various topics. Based on the similarity among those topics, we grouped them into three broad categories presented in Table \ref{task}: personal and well-being, activity and event planning, and professional development and learning. \textbf{Personal and well-being} includes tasks focused on self-improvement in physical or mental health, everyday scheduling, or interpersonal relationships, often recurring. \textbf{Activity and event planning} includes tasks related travel itineraries, route selection, and life events, typically requiring coordination of multiple factors such as preferences and budget. \textbf{Professional development and learning} includes tasks related to work, study, and research, often involving knowledge acquisition, strategic planning, or content generation. These topics reflect the broad applicability of AI-assisted planning across user groups and life contexts. 

\begin{table}
\caption{Task topics grouped in three main categories.}
\label{task}
\centering
\hspace*{-0.3cm}
\small
\begin{tabular}{l|l|l} \hline
\multicolumn{1}{c|}{\textbf{Category}} & \multicolumn{1}{c|}{\textbf{(Participant ID*) Task topics}} & \multicolumn{1}{c}{\textbf{Prompt source (count)}} \\ \hline
\multirow{6}{*}{\begin{tabular}[c]{@{}l@{}}Personal/\\ well-being\end{tabular}} & (E1) Anxiety management  & Template (3) \\ \cline{2-3}
 & (E7) Weekly fitness plan~~ & Self (12)~~ \\ \cline{2-3}
 & (R8) Girlfriend communication~~ & Template (4)~~ \\ \cline{2-3}
 & (R10) Weight loss plan~~ & Template (3)~~ \\ \cline{2-3}
 & (R11) Friends' avoidance~~ & Self (2) + Template (1) \\ \cline{2-3}
 & (R12) Study/workout schedule & Self (3) + Template (3) \\ \hline
\multirow{3}{*}{\begin{tabular}[c]{@{}l@{}}Activity/\\event \\planning\end{tabular}} & (E2) 14-day trip in Japan  & Template (3) \\ \cline{2-3}
 & (E4) Yellowstone trip~~ & Self (8) + Template (2) \\ \cline{2-3}
 & (R13) Birthday party planning & Template (3) \\ \hline
\multirow{5}{*}{\begin{tabular}[c]{@{}l@{}}Professional\\development/ \\learning \end{tabular}} & (E3) Health influencer strategy  & Self (15) \\ \cline{2-3}
 & (E5) Filmmaking preparation~~ & Self (8) + Template (2) \\ \cline{2-3}
 & (E6) Curiosity in later life~~ & Self (8) \\ \cline{2-3}
 & (R9) 3D model development~~ & Self (2) + Template (4) \\ \cline{2-3}
 & (R14) Microfinance learning & Self (2) + Template (1) \\ \hline
\multicolumn{3}{l}{\begin{footnotesize}*Recruited channel: email (E) and Reddit (R)\end{footnotesize}}
\end{tabular}
\end{table}

\subsubsection{General prompting behaviors}
Along with the task topics, Table \ref{task} also lists the source and number of prompts participants used. Prompts were either adapted from \textit{templates} provided by the researcher or written spontaneously by participant them\textit{selves}. Participants showed varying levels of engagement: some relied entirely on templates, others wrote all prompts themselves, and about half combined both, either starting with templates to explore AI capabilities or extending self-written conversations later. The number of prompts ranged from three to over ten, reflecting different engagement and confidence levels.

To summarize prompting behaviors, all prompts were categorized by their main functional role, such as initiating, refining, revising plans, providing context, or verifying information (Table~\ref{prompt}). Most prompts centered on providing background, initiating plans, or revising details, aligned with the template design. Self-generated prompts covered a broader range, extending to searching or verifying information, stating preference or constraints, generating ideas, and comparing options. Individual prompts can combine multiple functional roles. For instance, initiating a plan often incorporates providing context or specifying preferences. We classified each by its dominant function to capture general interaction patterns. Overall, participants engaged in iterative cycles of generation, refinement, and adaptation when planning with ChatGPT.

\begin{table*}
\caption{Prompt categories}
\label{prompt}
\centering
\small
\begin{tabular}{>{\hspace{0pt}}m{0.15\linewidth}|>{\hspace{0pt}}m{0.329\linewidth}|>{\hspace{0pt}}m{0.45\linewidth}} \hline
\multicolumn{1}{>{\centering\hspace{0pt}}m{0.15\linewidth}|}{\textbf{Category (count)}} & \multicolumn{1}{>{\centering\hspace{0pt}}m{0.329\linewidth}|}{\textbf{Description}} & \multicolumn{1}{>{\centering\arraybackslash\hspace{0pt}}m{0.45\linewidth}}{\textbf{Example prompts}} \\ \hline
Providing background/\par{}context (13) & The user describes their current situation, goal, or constraints (e.g., age, location, timeline). & \textit{``I'm 81…my goal in life is to learn more about how the world works. Do you have any advice for me?''} (E6) \\ \hline
Initiating a plan (11) & The user has a clear goal and wants the AI to generate a step-by-step plan, schedule, or strategy. & \textit{``Create a realistic weekly schedule that includes study, fitness, rest and side projects''} (R12) \\ \hline
Asking for details/\par{}refinement (24) & The user wants to dig deeper into or clarify parts of a generated plan or suggestion. & \textit{``your plan is very general, I have a lot of preferences and considerations. we can start with the Day 1 plan. Ask me more detail questions…''} (E2) \\ \hline
Revising or adjusting \par{}the plan (9) & The user has received an initial plan and now wants to adjust or optimize it. & \textit{``revise the exercise program to help me lose 2 lbs weekly if I eat 1500 calories daily…''} (E7) \\ \hline
Searching or verifying~\par{}information (15) & The user wants factual information or asks the AI to justify its answers. & \textit{``where did you get this information. do you have reliable sources or have you guessed at some answers…?''} (E7) \\ \hline
Stating preferences/\par{}constraints (7) & The user specifies likes/dislikes, conditions, or constraints related to the plan. & \textit{``Okay, I won't go to Rocky Mountain. I don't like long hiking''} (E4) \\ \hline
Generating ideas/\par{}content (6) & The user requests the AI to brainstorm or outline ideas, content, scripts, etc. & \textit{``What actions can I do to make her feel better''} (R8) \\ \hline
Comparing options (4) & The user is choosing between multiple options and wants help deciding. & \textit{``What would be more effective, a video of me walking in the park while talking, or my voice overlay on a music track?''} (E3) \\ \hline
\end{tabular}
\end{table*}

\subsubsection{Prompting in personal and well-being tasks}
Tasks in this category covered personal concerns such as fitness, stress, and relationships. Prompts typically began with a brief description of participants' situation or emotional state, such as \textit{``I'm preparing my PhD dissertation and feeling anxious; can you help me design an anxiety-management plan?''} (E1), or \textit{``I'm having trouble with my girlfriend; please help me make the relationship better.''} (R8). Participants used these openings to externalize difficulties and invite ChatGPT's general suggestions for improvement.

Participants then expanded on these prompts with contextual information, clarifying constraints or confirming details. For example, Participant E7 developed a weekly fitness routine through twelve prompts, providing background on age, schedule, and fitness level, then verifying information such as calorie and protein intake (\textit{``Where did you get this information?''}). This process illustrated how users alternated between generating plans and validating information. Across participants, prompting in this domain emphasized contextualization and confirmation rather than exhaustive specification. Users focused on checking whether ChatGPT's advice fit their situation and appeared realistic or trustworthy. Most treated ChatGPT's plans as provisional outlines that could later be adapted to real-life conditions.

\subsubsection{Prompting in activity and event planning}

Tasks in this category involved planning trips or events that required coordinating multiple factors such as timing, transportation, and budget. Compared with health-related tasks, these prompts emphasized specificity and feasibility. Participants aimed to turn ChatGPT's general outlines into actionable itineraries through repeated refinement.

For instance, Participant E2 began with a broad request for a Japan trip plan but then noted, \textit{``Your plan is very general; I have many preferences,''} prompting ChatGPT to ask clarifying questions before generating a day-by-day itinerary. Subsequent conversations refined flight routes, hotel safety, and budget, revealing a desire for control and contextual accuracy. Similarly, Participant E4, planning a Yellowstone trip, compared routes and schedules (\textit{``Would leaving from Denver take one more day?''}), personal preferences (\textit{``I don't like long hikes''}), and timing constraints (\textit{``My flight gets to Salt Lake City at 10 pm''}). These conversations highlight how participants negotiated between AI-generated generality and their own contextual knowledge. They refined plans until results felt realistic or until additional prompting no longer seemed productive.

\subsubsection{Prompting in professional development and learning}
Tasks in this category focused on career goals, academic learning, and creative or professional growth. Compared with other domains, these tasks were more open-ended and informationally uncertain, requiring verification and integration of new knowledge.

Participants such as E3, E5, and E6 treated ChatGPT less as a planner and more as a thinking partner, a conversational collaborator for exploring ideas and testing assumptions. Prompts combined background description, idea generation, and verification. For instance, participants asked ChatGPT to draft strategies or outlines, then iteratively refine content (\textit{``Make it sound more personal''}) or evaluate realism (\textit{``Would this income goal be realistic?''}). Other participants (R9, R14) used ChatGPT mainly for factual queries or practical information. Their interactions were short and task-oriented, focused on retrieving concrete information rather than reflection. Across these examples, professional and learning-oriented tasks ranged from exploratory to specific, depending on users' topic familiarity and goals. 

\subsection{Uncertainty in AI-assisted task planning}
For \textbf{RQ2}, we organized participants' reflections according to four interview questions related to uncertainty: (1) which aspects of planning felt uncertain; (2) how such uncertainty affected their process, confidence, and follow-through; (3) how AI reduced or amplified uncertainty; and (4) how planning without AI compared. Table~\ref{uncertainty} summarizes the key aspects and illustrative examples for each question. We elaborate on each question in the remainder of this section.

\begin{table*}
\caption{Uncertainty dimensions in AI-assisted task planning}
\label{uncertainty}
\centering
\small
\begin{tabular}{>{\hspace{0pt}}m{0.11\linewidth}|>{\hspace{0pt}}m{0.85\linewidth}} \hline
\textbf{Category} & \textbf{Key aspects and example quotes} \\ \hline
\multicolumn{2}{>{\hspace{0pt}}m{0.942\linewidth}}{\textit{Which part feels uncertain}} \\ \hline
\textbf{Task-inherent} & \textbf{Feasibility of plans} — \textit{``I wasn't sure if it was going to work out.''} (R10).  \textbf{Knowing where to start} — \textit{``I should ask but don't know what to ask.''} (E5).  \textbf{Balancing detail and flexibility} — \textit{``Too detailed plans make me anxious.''} (E1) \\
\textbf{AI-introduced} & \textbf{Contextual fit} — \textit{``I will arrive at night. It still planned as if daytime.''} (E4).  \textbf{Credibility} — \textit{``It creates citations that don't exist.''} (E6). \\ \hline
\multicolumn{2}{>{\hspace{0pt}}m{0.942\linewidth}}{\textit{Effect of uncertainty or uncertainty mitigation}} \\ \hline
\textbf{Emotion} & \textbf{Frustration} — \textit{``Uncertainty causes frustration… you feel like you can't achieve anything.''} (R9).  \textbf{Relief} (mitigated) — \textit{``If it's not like an assignment I must finish, I feel better.''} (E1).  \textbf{Growing confidence} (mitigated) — \textit{``The results were good… I gave it that confidence.''} (R10). \\
\textbf{Behavior} & \textbf{Iterative refinement} — \textit{``If it doesn't answer my question, I'll tweak it until I get what I want.''} (E7).  \textbf{Cross-checking} — \textit{``I have to fact check one by one.''} (E2); \textit{``Do more research and come back again.''} (R9).  \textbf{Simplifying plans} — \textit{``Sometimes I want to simplify this plan in two steps.''} (R12). \\ \hline
\multicolumn{2}{>{\hspace{0pt}}m{0.942\linewidth}}{\textit{How AI affects uncertainty}} \\ \hline
\textbf{Mitigate} & \textbf{Structuring goals} — \textit{``It reminded me to break things into smaller parts.''} (E1). \textbf{Encouraging reflection} — \textit{``He asked why I felt anxious… I started to think about it myself.''} (E1). \textbf{Providing new information} — \textit{``I never would have thought of these, you don't know what you don't know.''} (E3).  \textbf{Increasing confidence} — \textit{``I feel confident because it's guiding me.''} (R10). \\
\textbf{Exaggerate} & \textbf{Overreliance} — \textit{``You should have a basic understanding yourself before using ChatGPT.''} (E4).  \textbf{Lacking specificity} — \textit{``It gives disclosure but still generic.''} (R11). \\ \hline
\multicolumn{2}{>{\hspace{0pt}}m{0.942\linewidth}}{\textit{Uncertainty with non-AI methods}} \\ \hline
\textbf{Accessibility} & \textbf{Fact-checking} — \textit{``I can check from other websites or user reviews.''} (E2); \textit{``Books have structure and flow.''} (R9).  \textbf{Social discomfort} — \textit{``Hard to open up about personal issues… a friend may leak your information.''} (R8). \\
\textbf{Efficiency} & \textbf{Information overload} — \textit{``Online explanations are messy and overwhelming.''} (E4). \textbf{Time effort} — \textit{``Manual planning takes a lot of time and brain energy.''} (R12). \\ \hline
\end{tabular}
\end{table*}

\subsubsection{Which part feels uncertain}

Participants' reflections revealed two main sources of uncertainty: those inherent to the task itself and those introduced by the AI system.
Task-inherent uncertainty included doubts about the feasibility of plans, uncertainty about where to start, and tension between detail and flexibility. Several participants were unsure whether their goals were realistic or sustainable in practice: \textit{``I wasn't sure if it was going to work out''} (R10), or expressed confusion about how to begin a complex plan: \textit{``I should ask but don't know what to ask''} (E5). Others found that excessive structure could increase anxiety rather than reduce it, noting that \textit{``too detailed plans make me anxious''} (E1).

AI-introduced uncertainty, in contrast, stemmed from how ChatGPT generated and presented information. Participants questioned the contextual fit of its suggestions: \textit{``I will arrive at night. It still planned as if daytime''} (E4), and the credibility of its content: \textit{``It creates citations that don't exist''} (E6). These accounts highlight that uncertainty emerged both from the nature of the planning task itself and from limitations in how AI interpreted and contextualized user inputs.

\subsubsection{Effect of uncertainty or uncertainty mitigation}

Uncertainty shaped both the emotional and behavioral dimensions of participants' interactions with ChatGPT.
Emotionally, it elicited a mix of frustration, satisfaction, and growing confidence. Some participants described discouragement when facing incomplete or inconsistent outputs: \textit{``Uncertainty causes frustration… you feel like you can't achieve anything''} (R9), while others felt more at ease when plans were flexible and non-obligatory: \textit{``If it's not like an assignment I must finish, I feel better''} (E1). Over time, positive outcomes helped transform skepticism into trust when uncertainty reduced: \textit{``The results were good… I gave it that confidence''} (R10).

Behaviorally, uncertainty led to iterative refinement, cross-checking, and simplifying plans. When answers were vague, participants adjusted their prompts: \textit{``If it doesn't answer my question, I'll tweak it until I get what I want''} (E7), or verified information externally: \textit{``I have to fact check one by one''} (E2). Some simplified long or complex plans into shorter, clearer steps: \textit{``Sometimes I want to simplify this plan in two steps''} (R12). In this way, uncertainty became a driver for clarification and learning rather than a negative experience.

\subsubsection{How AI affects uncertainty}

Participants described ChatGPT as both mitigating and amplifying uncertainty.
In many cases, AI helped structure goals, encourage reflection, provide new information, and increase confidence. For example, one participant noted that ChatGPT \textit{``reminded me to break things into smaller parts''} (E1), while another explained, \textit{``He asked why I felt anxious… I started to think about it myself''} (E1). Others appreciated how the system surfaced novel ideas: \textit{``I never would have thought of these, you don't know what you don't know''} (E3), or how clear guidance boosted motivation: \textit{``I feel confident because it's guiding me''} (R10). Through such mechanisms, ChatGPT reduced uncertainty by offering structure and prompting self-awareness.

At the same time, AI sometimes exaggerated uncertainty by fostering overreliance or producing insufficiently specific advice. Some participants cautioned that using ChatGPT effectively required their own background knowledge: \textit{``You should have a basic understanding yourself before using ChatGPT''} (E4), while others found its responses too generic to act on: \textit{``It gives disclosure but still generic''} (R11). Thus, while ChatGPT often clarified what to do, it could also blur boundaries between confidence and dependency.

\subsubsection{Uncertainty with non-AI methods}

When comparing AI-assisted planning with traditional approaches, participants highlighted differences in accessibility and efficiency.
Regarding accessibility, non-AI methods such as human experts or reference materials supported fact-checking and provided structured knowledge: \textit{``I can check from other websites or user reviews''} (E2); \textit{``Books have structure and flow''} (R9), but they also involved social discomfort, especially for personal topics: \textit{``Hard to open up about personal issues… a friend may leak your information''} (R8).

In terms of efficiency, participants emphasized the challenges of information overload and time effort. Searching online was described as \textit{``messy and overwhelming''} (E4), and manual planning as \textit{``taking a lot of brain energy''} (R12). Compared with these experiences, ChatGPT was valued for its immediacy and cognitive relief, offering concise summaries and a private, always-available space for reflection. Together, these accounts suggest that while non-AI methods reduce uncertainty through credibility and expertise, AI expands accessibility and efficiency, shifting how people define what is feasible or worth planning.

\subsection{Overall usefulness and limitations}
For \textbf{RQ3}, Table \ref{impression} presents results of participants impression and feedback on the plan's usefulness and limitations. Participants generally described ChatGPT as both useful and limited in supporting their planning processes. Across interviews, two recurring strengths stood out: its ability to structure complex goals into manageable steps, and its conversational tone that made planning feel less solitary. At the same time, participants were candid about the system's limitations, particularly its lack of personalization, contextual realism, and adaptive follow-up. These perceptions reveal how users positioned ChatGPT not only as a task assistant but as a reflective partner that helped them externalize thinking, negotiate uncertainty, and stay motivated.

\begin{table*}
\caption{Participants' impression on task planning using ChatGPT}
\label{impression}
\centering
\small
\begin{tabular}{l|l|l} \hline
\multicolumn{1}{c|}{\textbf{Theme}} & \multicolumn{1}{c|}{\textbf{Description}} & \multicolumn{1}{c}{\textbf{Example quotes}} \\ \hline
\begin{tabular}[c]{@{}l@{}}\textbf{Scaffolding and }\\\textbf{motivational support}\end{tabular} & \begin{tabular}[c]{@{}l@{}}Helped users start, decompose \\goals, and feel progress\end{tabular} & \begin{tabular}[c]{@{}l@{}}\textit{``Dividing big goals into small tasks gave me a sense of control.''} (E1) / \textit{``I can start} \\\textit{putting parts together to a fairly comprehensive social media campaign.''} (E3)\end{tabular} \\ \hline
\begin{tabular}[c]{@{}l@{}}\textbf{Reflective and }\\\textbf{conversational value}\end{tabular} & \begin{tabular}[c]{@{}l@{}}Served as sounding board; confirmed \\instincts or pushed reflection\end{tabular} & \begin{tabular}[c]{@{}l@{}}\textit{``It pushes my thinking.''} (E6) /~\textit{``It confirms I can do this.''} (E1) / \textit{``It takes that} \\\textit{position of you... like another human being on the other side.''} (R8)~\end{tabular} \\ \hline
\begin{tabular}[c]{@{}l@{}}\textbf{Personalization and }\\\textbf{realism limits}\end{tabular} & \begin{tabular}[c]{@{}l@{}}Plans too general or idealized; \\missed context details\end{tabular} & \begin{tabular}[c]{@{}l@{}}\textit{``It never asks me how many people.''} (E4) /~\textit{``It's just an outline… not real-life }\\\textit{enough.''} (R12) /~\textit{``Plans need to be realistic about my free time.''}~(R12)\end{tabular} \\ \hline
\begin{tabular}[c]{@{}l@{}}\textbf{Trust and control }\\\textbf{expectations}\end{tabular} & \begin{tabular}[c]{@{}l@{}}Users verified info; wanted adaptive \\memory, tone control, and transparency\end{tabular} & \begin{tabular}[c]{@{}l@{}}\textit{``I would check Mayo Clinic for reliability.''} (E7) / \textit{``I can't apply the code directly.''} (R9) \\ \textit{``If it didn't compliment people so much, it would be more believable.''} (E3)\end{tabular} \\ \hline
\end{tabular}
\end{table*}

\subsubsection{Perceived usefulness: structure, motivation, and reflection}

Participants widely appreciated ChatGPT's ability to transform abstract goals into concrete steps. It offered structure, measurable targets, and a sense of progress that supported motivation: \textit{``Dividing big goals into small tasks gave me a sense of control''} (E1). Beyond practical organization, many described ChatGPT as a conversational partner that stimulated reflection and emotional reassurance. \textit{``It pushes my thinking''} (E6) and \textit{``It confirms that I can do this''} (E1) exemplify how participants perceived the system as both supportive and reflective, enabling them to reason through decisions and sustain engagement over time.

\subsubsection{Perceived limitations: personalization, realism, and credibility}

Despite these benefits, participants frequently emphasized the generic and idealized nature of AI-generated plans. Even when detailed information was provided, outputs often ignored personal constraints or contextual nuances: \textit{``It never asks me how many people, the plan will be totally different''} (E2). Others criticized the lack of realism: \textit{``It's just an outline… not real-life enough''} (R12), and expressed frustration with having to repeatedly adjust prompts. Concerns about credibility were also common. Several participants questioned the factual grounding of responses, choosing to verify them through external sources such as medical or educational websites: \textit{``I would check Mayo Clinic for reliability''} (E7).

\subsubsection{Expectations for improvement and adaptive control}

Participants' reflections on these limitations translated into clear expectations for future AI-assisted planning tools. They wanted systems that could remember personal context, adapt to progress, and explain the basis of their suggestions. Many also desired finer control over tone and style, ranging from supportive to more critical feedback, rather than default encouragement. As one participant summarized, \textit{``If it didn't compliment people so much, it would be more believable''} (E3), suggesting that personalization and situational awareness were viewed as essential for genuine assistance.

Overall, participants recognized ChatGPT's dual role as a planning scaffold and a reflective partner. Its ability to structure goals and maintain a human-like tone made it valuable for initiating and sustaining planning. Yet its limitations in personalization, realism, and credibility constrained users' trust and reliance. These mixed impressions mirror a broader tension throughout the study: ChatGPT helps users think and plan, but it also transfers part of the cognitive effort back to them. In this sense, prompting becomes an ongoing act of negotiation, between structure and flexibility, automation and agency, and guidance and trust.

\section{Discussion}
Across the three RQs, we found that participants used ChatGPT for diverse long-term planning tasks and developed prompting strategies to set goals, decompose steps, and refine outputs (RQ1). Yet planning was marked by significant uncertainty: users were unsure how to start, which options were realistic, and whether AI advice could be trusted or followed through (RQ2). While ChatGPT was valued for scaffolding, idea generation, and encouragement, participants also highlighted its lack of personalization, realism, adaptability, and credibility. They expressed a desire for systems that could remember context, dynamically adjust, and provide transparent, trustworthy guidance (RQ3). Overall, ChatGPT functioned as a helpful scaffolder and reflective partner, but its current limitations constrained its effectiveness for long-term task planning.

\vspace{-0.3cm}
\subsection{The diversity of planning goals and needs}

We observed three main task categories in this study, including personal and well-being, activity and event planning, and professional development and learning. Participants adopted specific prompting behaviors aligned with the nature of their tasks, showing how people adapt their interaction style with AI to different planning needs. By examining these prompting patterns together with the inherent characteristics of each task type, we find that the three categories reflect distinct demands and expectations for AI assistance.

For personal and well-being tasks, participants preferred open-ended and flexible prompting and often resisted detailed or prescriptive plans. These goals, such as maintaining health, managing stress, or improving relationships, were integrated into everyday life and could be easily influenced by changing routines or emotions \citep{kuhlthau_inside_1991, abbott_understanding_2005}. In contrast, activity and event planning tasks, like travel or event organization, had concrete deadlines and interdependent factors such as timing, logistics, and budget \citep{bystrom_conceptual_2005, xie_travelplanner_2024}. Participants sought specific and executable plans, iteratively refining ChatGPT's suggestions until the results appeared realistic and complete. Meanwhile, professional development and learning tasks were more exploratory and information-driven \citep{vakkari_task-based_2003, shah_taking_2023}. Users engaged ChatGPT to structure abstract ambitions, explore new domains, or build conceptual roadmaps rather than fixed action steps. Together, these patterns show how the temporal scope and uncertainty of a goal shape users' prompting strategies and expectations for AI assistance: goals with recurring tasks favor adaptive planning and general guidance; bounded tasks call for precise and front-loaded planning; and open-ended, knowledge-oriented goals rely on conceptual scaffolding and reflection \citep{norman_attention_1986, liao_scaffolding_2024}.

\subsection{Uncertainties in planning with AI}

Across task types, participants described uncertainty arising from two main sources: the inherent ambiguity of the task itself and the limitations introduced by the AI system. Task-inherent uncertainty included doubts about feasibility, knowing where to start, or balancing detail and flexibility, whereas AI-introduced uncertainty stemmed from issues such as lack of contextual fit, limited personalization, or questionable credibility. These two sources shaped participants' experiences differently. AI-introduced uncertainty primarily influenced how users perceived the generated plan, whether it seemed realistic, trustworthy, or personally relevant, while task-inherent uncertainty was tied to how they imagined carrying out the plan in the real world.

Taken together, these patterns suggest that users experienced uncertainty not only in different forms but also at different temporal stages of the planning process. AI-introduced uncertainty tended to affect the present interaction with ChatGPT, while task-inherent uncertainty extended into the future enactment of the plan. This temporal distinction resonates with \textit{construal level theory} \citep{trope_construal-level_2010}, which posits that people think about distant future goals in abstract terms and near-term actions in concrete terms. In our data, participants' reflections mirrored this temporal gradient: uncertainty can shift from abstract questions about how to plan toward more situated concerns about how to act \citep{gollwitzer_implementation_1999}.

To interpret this distinction, we can tentatively describe two interrelated dimensions of uncertainty: \textit{planning uncertainty} (i.e., ambiguity about how to define and structure a plan) and \textit{doing uncertainty} (i.e., apprehension that arises when imagining the actual execution of that plan). ChatGPT was effective at easing the former by providing structure, generating ideas, and clarifying options, but less capable of addressing the latter, which was emotional and context-dependent. This distinction helps explain why participants' preferences for plan specificity diverged across task types: those engaged in exploratory goals (e.g., professional learning) were comfortable with general plans that preserved flexibility; those managing evolving conditions (e.g., health or relationships) found detailed plans restrictive or unrealistic; and those pursuing bounded, short-term tasks (e.g., travel or events) desired specific, executable guidance.
\vspace{-0.2cm}
\subsection{Design implications}
Our findings suggest that AI plays two complementary roles in long-term planning: as a cognitive scaffold and as a reflective partner. As a scaffold, ChatGPT helps users externalize and structure complex goals, break them into actionable steps, and maintain a sense of progress. Scaffolding can support cognitive organization and task success \citep{vygotskij_mind_1981, camara_searching_2021, zhou_exploring_2025}. As a reflective partner, it sustains engagement by offering conversational feedback, emotional reassurance, and adaptive suggestions that help users manage uncertainty and motivation over time \citep{miceli_anxiety_2005}. However, these same strengths also expose important limitations. ChatGPT's encouraging tone and success-oriented framing often led to overly optimistic or flattering responses, depicting idealized futures while overlooking potential setbacks, contextual constraints, or emotional challenges \citep{zhang_supervised_2025, sivaprasad_theory_2025}. By consistently confirming users' goals and emphasizing positive outcomes, the system sometimes obscured signals of possible failure and fostered overconfidence in the feasibility of its plans \citep{carro_flattering_2024, cheng_elephant_2025}. This dynamic highlights a central design tension: while scaffolding and encouragement can enhance motivation and perceived clarity, they may also mask the inherent uncertainty of real-world action.

Taken together, these findings point to several design implications for AI-assisted planning tools. Systems should support personalization and user modeling, capturing context progressively rather than requiring exhaustive upfront input. They should enhance explainability and transparency, providing sources, confidence levels, and explicit reasoning pathways. To better manage uncertainty, AI should move beyond success-oriented planning and incorporate mechanisms for simulating obstacles, surfacing common points of failure, and suggesting contingency plans. Moreover, tools should explicitly support iterative and adaptive workflows, remembering user preferences across sessions and dynamically adjusting as users report progress or difficulties. Motivational support also remains important, but it should be configurable, allowing users to switch between encouragement and flattery. Finally, privacy and sensitivity issues must be carefully handled, especially as personalization and cross-session memory become central to long-term planning.

\subsection{Limitations and future directions}
This study has several limitations that should be considered when interpreting the findings. As an exploratory qualitative study with a small sample, the results reflect the perspectives of participants who were already motivated to experiment with AI, which may limit generalizability. The study also examined planning scenarios as examples and relied on self-reported reflections rather than observing how AI-assisted plans in real practice. Moreover, the findings are tied to a specific model at a particular stage of technological development, and users' experiences may shift as AI systems become more adaptive and integrated into daily workflows. Nevertheless, the study offers a grounded understanding of how people engage with AI for long-term life task planning, revealing patterns of use, forms of uncertainty, and the role of AI. These insights inform future directions for research and design.

Future work could extend this direction by examining more diverse user groups and longitudinal settings, capturing how AI-assisted plans unfold over time and how uncertainty changes across different task horizons. Design-oriented studies may also explore systems that better account for failure and change, support shared or multi-task planning, and incorporate adaptive memory, transparency, and feedback mechanisms. More broadly, advancing toward resilient and trustworthy planning support will require positioning AI not merely as a plan generator but as a collaborative and context-aware partner that evolves alongside users' goals.

\section{Conclusion}
This study examined how people use ChatGPT for long-term life task planning, a domain marked by complexity and uncertainty. Through a mixed analysis of participants' prompts and interview reflections, we found that ChatGPT provided valuable scaffolding, idea generation, and encouragement, but struggled with personalization, realism, adaptability, and credibility. It reveals both the promise and the limits of current AI systems. Our findings contribute to understanding AI as a collaborative planning partner and highlight design opportunities for more adaptive, trustworthy, and failure-aware planning tools that better support human uncertainty in real-world contexts.

\bibliographystyle{ACM-Reference-Format}
\bibliography{sample-base}

\appendix

\end{document}